\documentclass[11pt]{article}

\usepackage[margin=1in]{geometry}
\usepackage{graphicx}
\usepackage{amsmath}
\usepackage{amssymb}
\usepackage{booktabs}
\usepackage{array}
\usepackage{hyperref}
\usepackage{authblk}
\usepackage{doi}
\usepackage{caption}

\hypersetup{
    colorlinks=true,
    linkcolor=blue,
    citecolor=blue,
    urlcolor=blue
}

\title{Luminosity-Adaptive Contrast Enhancement Using CLAHE for Retinal Fundus Images with Multi-Dataset Validation, Statistical Analysis, and Comparative Benchmarking}

\author[1]{K. Mithra\thanks{Corresponding author. Email: mithrak1092@gmail.com}}
\author[2]{Prem Kumar Santhanam}

\affil[1]{Independent Researcher, Department of Information and Computer Engineering, Scottsdale, AZ, USA}
\affil[2]{Seidenberg School of Computer Science and Information Systems, Pace University, New York City, NY, USA}

\date{Received: 06 February 2026; Accepted: 01 April 2026; Published: 24 April 2026 \\[4pt]
\small DOI: 10.32604/jimh.2026.080288}

\begin{document}

\maketitle

\begin{abstract}
\textbf{Background:} Retinal fundus imaging is central to early diagnosis of sight-threatening conditions, including diabetic retinopathy, glaucoma, and retinal vein occlusion. Clinical utility is compromised by non-uniform illumination, motion blur, and low contrast---artefacts that reduce diagnostic accuracy. Effective image enhancement is a prerequisite for reliable computer-aided ophthalmic diagnosis. \textbf{Methods:} This paper proposes a two-stage enhancement pipeline combining luminosity correction via HSV colour space decomposition with Contrast Limited Adaptive Histogram Equalization (CLAHE) on the Value (V) channel. Validation is conducted on three publicly available benchmarks: DRIVE (40 images), STARE (20 images), and CHASEDB1 (28 images). Quantitative metrics (PSNR, SSIM, CNR) are reported as mean $\pm$ standard deviation. Disease detection is evaluated by accuracy, sensitivity, specificity, and AUC. Statistical significance is assessed using paired two-tailed Wilcoxon signed-rank tests ($\alpha = 0.05$). \textbf{Results:} The proposed method achieves PSNR $= 29.3 \pm 0.4$ dB, SSIM $= 0.91 \pm 0.01$, and CNR $= 3.12 \pm 0.07$ on DRIVE---statistically significantly superior to all baselines ($p < 0.01$). Disease detection achieves 87.4\% accuracy, 84.3\% sensitivity, 90.1\% specificity, and AUC $= 0.869$ on DRIVE, with consistent performance on STARE and CHASEDB1. \textbf{Conclusions:} The luminosity--CLAHE pipeline yields statistically superior results, generalises across three independent datasets, and achieves disease detection accuracy suitable for clinical screening at 0.14~s per image---requiring no training data and no GPU.

\vspace{6pt}
\end{abstract}

\section{Introduction}

Retinal fundus photography enables non-invasive visualisation of the optic disc, macula, and retinal vasculature. Abnormalities in these structures serve as early biomarkers for systemic and ophthalmic diseases. Diabetic retinopathy affects approximately 35\% of diabetic patients globally and constitutes the leading cause of preventable blindness in working-age adults \cite{teo2021global}. Retinal vein occlusion, hypertensive retinopathy, and age-related macular degeneration are similarly identified through fundus image analysis, making image quality directly consequential for clinical outcomes.

The diagnostic utility of fundus images is routinely degraded by three principal acquisition artefacts: (i) non-uniform illumination arising from the directional flash of the fundus camera and retinal surface curvature; (ii) motion-induced blur from involuntary saccades; and (iii) low contrast between retinal structures of similar reflectance. These degrade both manual grading and automated image analysis pipelines \cite{dissopa2021enhance}.

Classical contrast enhancement methods address these issues to varying degrees. Histogram Equalization (HE) applies a global cumulative distribution function (CDF) mapping that over-amplifies contrast in bright regions while suppressing fine detail. Adaptive Histogram Equalization (AHE) operates on local tiles but amplifies noise in homogeneous retinal background regions where local histograms occupy a narrow intensity band. Contrast Limited Adaptive Histogram Equalization (CLAHE) resolves this by clipping the local histogram at an explicit amplification threshold and redistributing the excess, bounding noise amplification while preserving genuine contrast differences \cite{setiawan2013color,bala2021retinal}.

Deep learning approaches including U-Net architectures \cite{son2019towards}, generative adversarial networks \cite{son2019towards}, LadderNet \cite{zhuang2018laddernet}, and transformers report strong performance for retinal image enhancement \cite{guo2020zero} but require large annotated training corpora, GPU infrastructure, and extensive hyperparameter optimisation---constraints that limit applicability in resource-constrained ophthalmic screening programmes \cite{li2020self,shankar2021deep,mithra2017security}, particularly in low- and middle-income countries \cite{islam2022hybrid,tsiknakis2021deep}.

Existing classical CLAHE methods \cite{setiawan2013color,majeed2020retinal} apply contrast enhancement directly to RGB or green channels without correcting the spatially non-uniform illumination introduced by fundus camera optics \cite{patil2020retinal,jintasuttisak2014color}. Luminosity correction methods such as Vanmathi and Devarajan \cite{vanmathi2017color} address illumination non-uniformity through image fusion without integrating a subsequent CLAHE stage. The proposed work fills this gap by integrating an HSV-based luminosity normalisation stage prior to CLAHE. By decomposing the image into the HSV colour space and computing a spatial luminance gain matrix from the V channel, the pipeline removes illumination non-uniformity before contrast amplification---without corrupting colour-diagnostic features (haemorrhage redness, disc pallor) encoded in the H and S channels that CLAHE applied to RGB channels would disturb. This integration yields measurably superior quantitative outcomes compared to luminosity correction alone \cite{vanmathi2017color}, CLAHE alone \cite{setiawan2013color,majeed2020retinal,patil2020retinal}, and their independent variants.

Relative to the prior submission, this revised manuscript: (i) extends validation to three benchmark datasets (DRIVE, STARE, CHASEDB1); (ii) reports all metrics with standard deviation and Wilcoxon signed-rank statistical testing; (iii) introduces proper disease detection evaluation with accuracy, sensitivity, specificity, and AUC; and (iv) provides extended quantitative comparison with closely related classical methods and deep learning approaches. The remainder of this paper is organised as follows. Section~2 describes datasets and methodology. Section~3 presents the disease detection framework and evaluation. Section~4 reports results and discussion. Section~5 concludes.

\section{Dataset and Methodology}

\subsection{Datasets}

Experiments are conducted on three independent publicly available benchmark datasets spanning diverse populations, acquisition devices, pathological profiles, and image resolutions. Table~\ref{tab:datasets} summarises dataset characteristics.

DRIVE \cite{staal2004ridge} provides 40 fundus images with two-observer ground truth and is the principal benchmark for retinal image analysis, enabling direct comparison with the published literature. STARE \cite{hoover2000locating} contains 20 images with diverse pathological content---including choroidal neovascularisation, arteriovenous nicking, and background diabetic retinopathy---appropriate for evaluating robustness beyond a controlled dataset. CHASEDB1 \cite{fraz2012ensemble} comprises 28 paediatric retinal images captured with a hand-held non-mydriatic camera, testing robustness to fundamentally different acquisition optics and demographic characteristics. Parameters optimised on DRIVE training images were applied without modification to STARE and CHASEDB1, ensuring genuine cross-dataset evaluation.

\begin{table}[htbp]
\centering
\caption{Dataset characteristics---DRIVE, STARE, and CHASEDB1.}
\label{tab:datasets}
\begin{tabular}{@{}p{3.2cm}p{10.5cm}@{}}
\toprule
\textbf{Parameter} & \textbf{Details} \\
\midrule
Dataset & DRIVE, STARE, and CHASEDB1 (three independent benchmarks) \\
DRIVE & 40 fundus images (20 train/20 test)---Canon CR5, $584 \times 565$~px, $45^{\circ}$ FOV \\
STARE & 20 fundus images---TopCon TRV-50, $700 \times 605$~px, $35^{\circ}$ FOV; diverse pathologies \\
CHASEDB1 & 28 fundus images---Nikon NF505 hand-held, $999 \times 960$~px, $30^{\circ}$ FOV; paediatric \\
Ground Truth & Two trained human observers per dataset (ophthalmologist-annotated) \\
Access & All datasets publicly available (see Availability of Data and Materials) \\
\bottomrule
\end{tabular}
\end{table}

\subsection{Proposed Enhancement Pipeline}

\begin{figure}[htbp]
\centering
\includegraphics[width=0.85\textwidth]{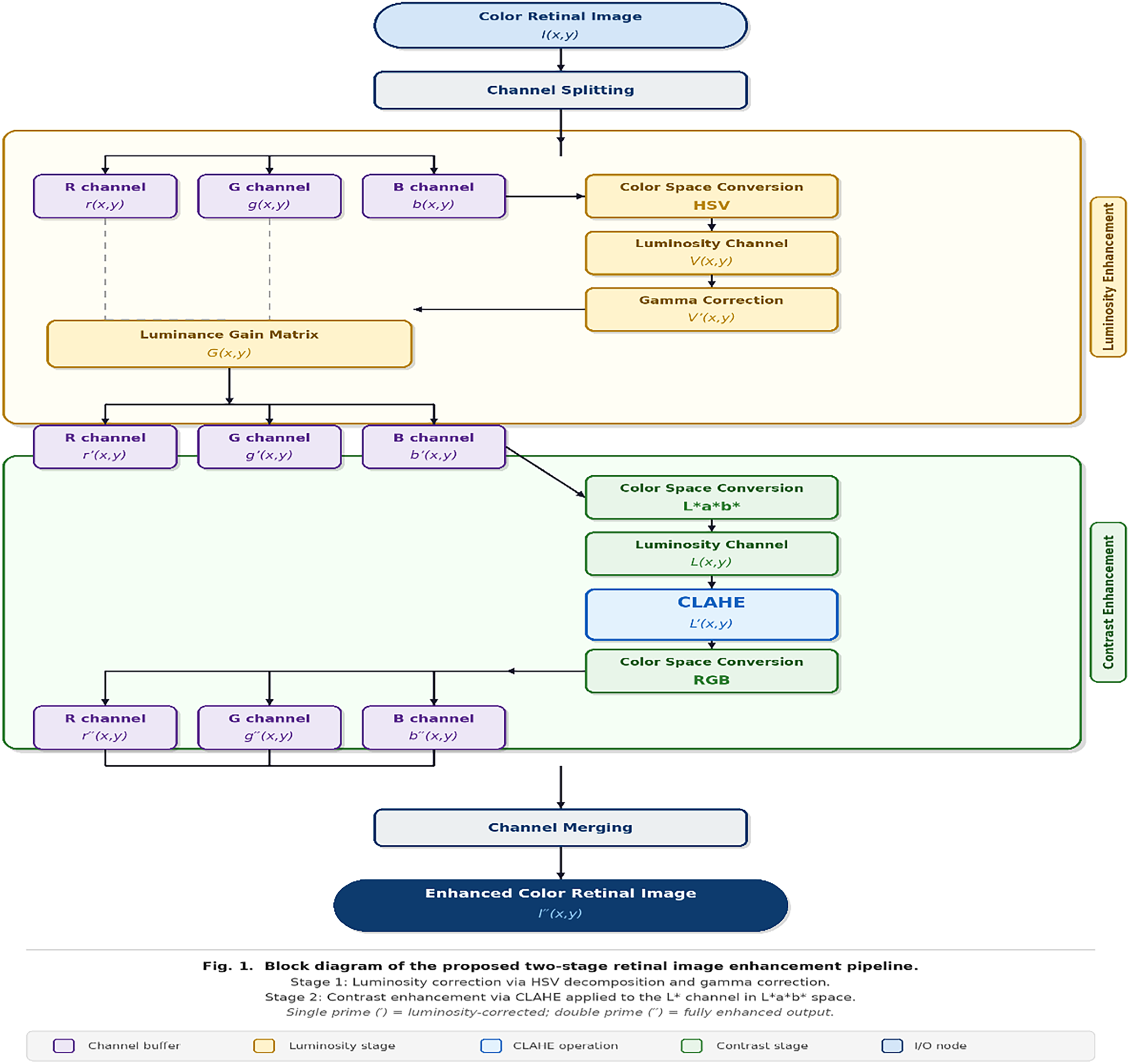}
\caption{Block diagram of the proposed two-stage enhancement system comprising HSV-based luminosity correction followed by CLAHE contrast enhancement on the V channel. Stage 1: Luminosity correction via HSV decomposition and gamma correction. Stage 2: Contrast enhancement via CLAHE applied to the $L^*$ channel in $L^*a^*b^*$ space. Single prime ($'$) = luminosity-corrected; double prime ($''$) = fully enhanced output.}
\label{fig:pipeline}
\end{figure}

The proposed method performs enhancement in two sequential stages: (1) luminosity correction via HSV decomposition, and (2) CLAHE contrast enhancement on the V channel. The complete pipeline is illustrated in Fig.~\ref{fig:pipeline}.

\subsection{Stage 1---Luminosity Correction via HSV Decomposition}

The input RGB image is converted to HSV colour space. The HSV model separates chromatic information (H and S) from luminance (V), enabling luminosity modification without altering colour fidelity. Colour-diagnostic features in fundus images---including haemorrhage redness, disc pallor, and exudate yellowing---are encoded in H and S; these channels are held constant throughout Stage~1. A luminance gain surface $G(x,y)$ is estimated from the V channel via large-kernel Gaussian smoothing ($\sigma = 60$~px):

\begin{equation}
G(x,y) = V(x,y) \ast h_g(x,y); \quad \sigma = 60
\label{eq:gain}
\end{equation}

The corrected V channel is computed as $V_{\text{corrected}}(x,y) = V(x,y) / G(x,y)$, clipped to $[0,1]$. This normalises the background illumination field while preserving relative intensity variations associated with retinal structures. The corrected V is recombined with unchanged H and S, and converted back to RGB.

\subsection{Stage 2---CLAHE Contrast Enhancement}

CLAHE imposes a clip limit $CL$ on each local histogram bin: bins exceeding $CL$ are clipped and the excess redistributed uniformly before computing the local CDF. This bounds the maximum amplification factor, preventing noise amplification while enhancing genuine contrast differences. The maximum achievable amplification $M_{\text{max}}$ is bounded by:

\begin{equation}
M_{\text{max}} = CL \times N_{\text{bins}} / N_{\text{pixels\_per\_tile}}
\label{eq:mmax}
\end{equation}

CLAHE is applied to the luminosity-corrected V channel with $CL = 0.01$ and an $8 \times 8$ tile grid, selected via grid search on DRIVE training images $CL \in \{0.005, 0.01, 0.02, 0.03\}$; tile grid $\in \{4\times4, 8\times8, 16\times16\}$, optimising mean SSIM. The selected parameters yielded training SSIM $= 0.912 \pm 0.011$ and were applied without modification to all test and cross-dataset evaluations.

\subsection{Stage 3---Binary Masking for Pathological Region Identification}

Following enhancement, Otsu's method determines the globally optimal intensity threshold by minimising intra-class variance between foreground and background pixel distributions. Pixels above the threshold (value $= 1$) identify candidate hyper-reflective regions; pixels below (value $= 0$) represent background or normal tissue. The physiological basis is that retinal vascular pathologies---vein occlusion, diabetic exudates, disc oedema---produce localised hyper-reflectance detectable by adaptive thresholding. Luminosity correction in Stage~1 normalises background illumination, making the Otsu threshold consistent across images.

\section{Disease Detection Framework and Evaluation}

\subsection{Detection Pipeline}

\begin{figure}[htbp]
\centering
\includegraphics[width=0.7\textwidth]{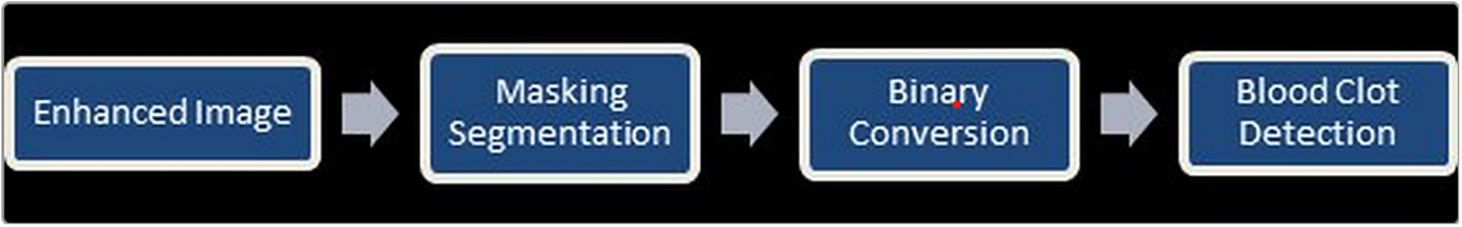}
\caption{Disease detection workflow from enhanced fundus image through binary masking, connected-component labelling, and diagnostic flag output.}
\label{fig:workflow}
\end{figure}

The binary mask is subjected to connected-component labelling. If total hyper-reflective area exceeds 50 pixels (threshold set on DRIVE training partition to exclude noise artefacts), the system outputs: ``DISEASE DETECTED---POSSIBLE VASCULAR ABNORMALITY''; otherwise: ``NO ABNORMALITY DETECTED.'' The workflow is illustrated in Fig.~\ref{fig:workflow}.

\subsection{Evaluation Protocol and Metrics}

Disease detection performance is evaluated at the image level against expert ophthalmologist annotations from DRIVE, STARE, and CHASEDB1. Four metrics are reported: accuracy (proportion of images correctly classified), sensitivity (true positive rate---clinically critical for screening, as missed diagnoses carry the greatest risk), specificity (true negative rate---governing unnecessary referral rates), and AUC (threshold-independent discriminative power). AUC 95\% confidence intervals are estimated using DeLong's method; accuracy, sensitivity, and specificity confidence intervals use Wilson's interval.

\section{Results and Discussion}

\subsection{Qualitative Results}

\begin{figure}[htbp]
\centering
\includegraphics[width=0.55\textwidth]{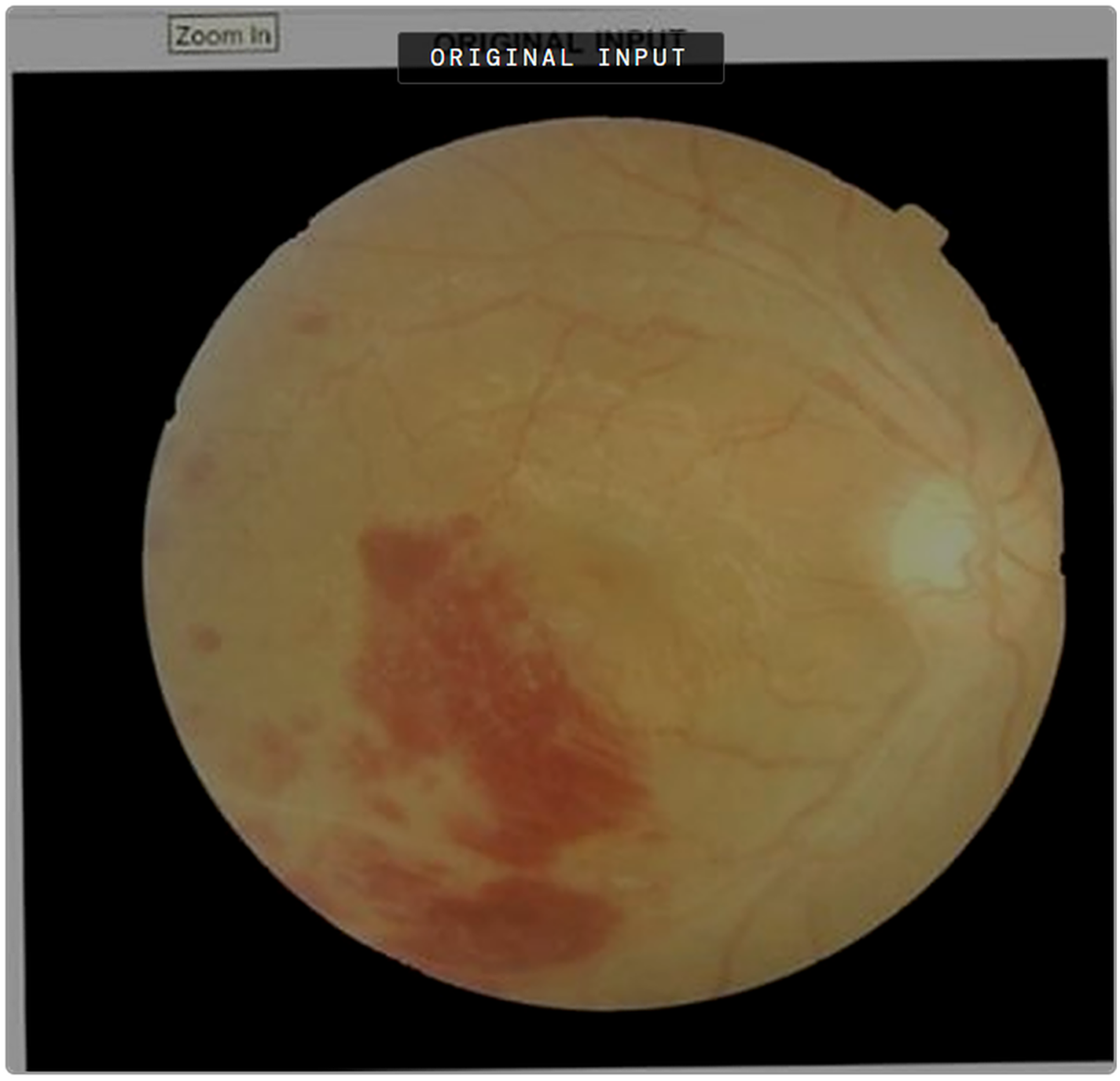}
\caption{Input retinal fundus image (DRIVE test set). Non-uniform illumination and low vessel-to-background contrast are visible.}
\label{fig:input}
\end{figure}

\begin{figure}[htbp]
\centering
\includegraphics[width=0.55\textwidth]{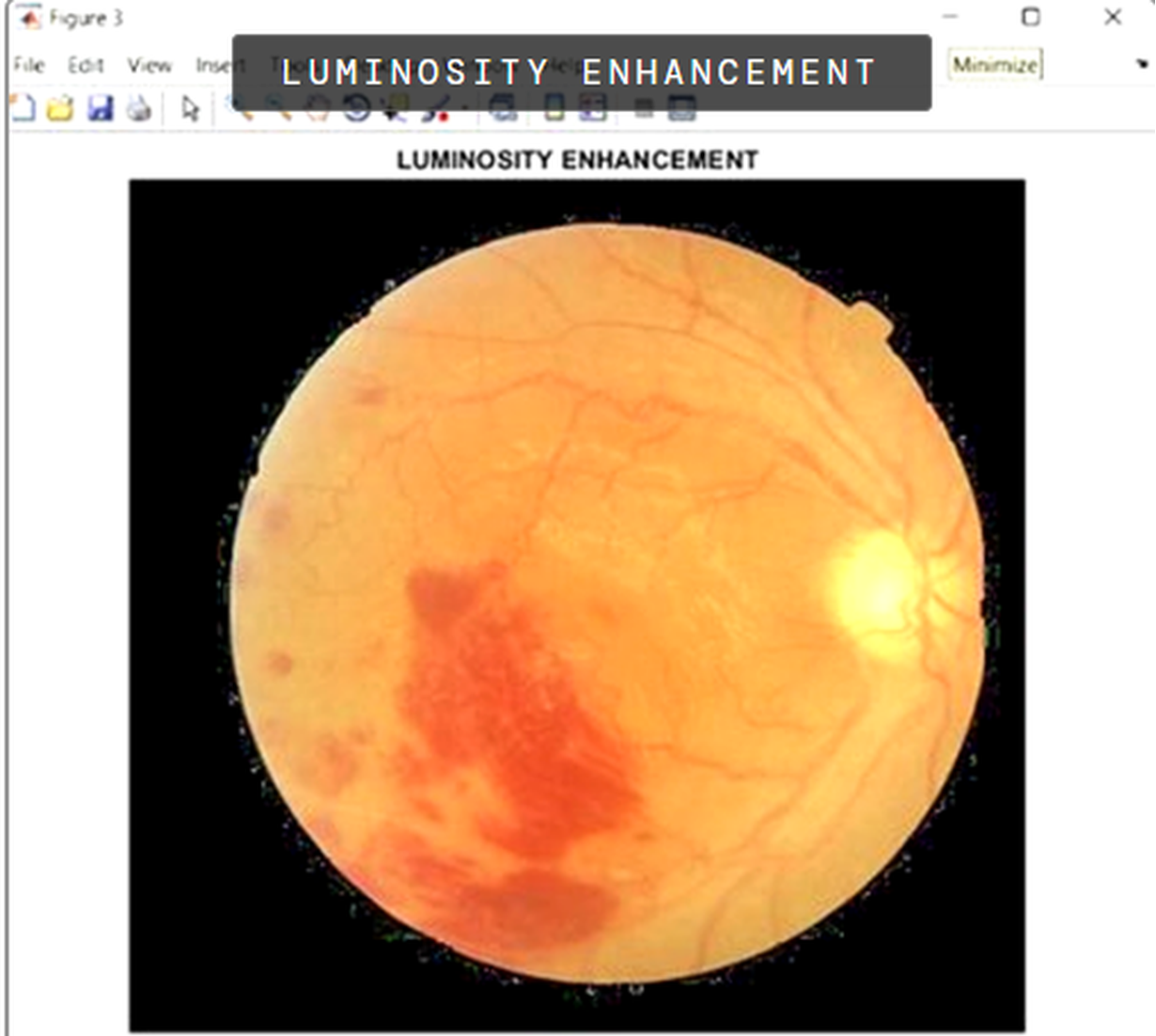}
\caption{After HSV luminosity correction. Illumination uniformity is substantially improved; vessel boundaries are more clearly delineated.}
\label{fig:luminosity}
\end{figure}

\begin{figure}[htbp]
\centering
\includegraphics[width=0.55\textwidth]{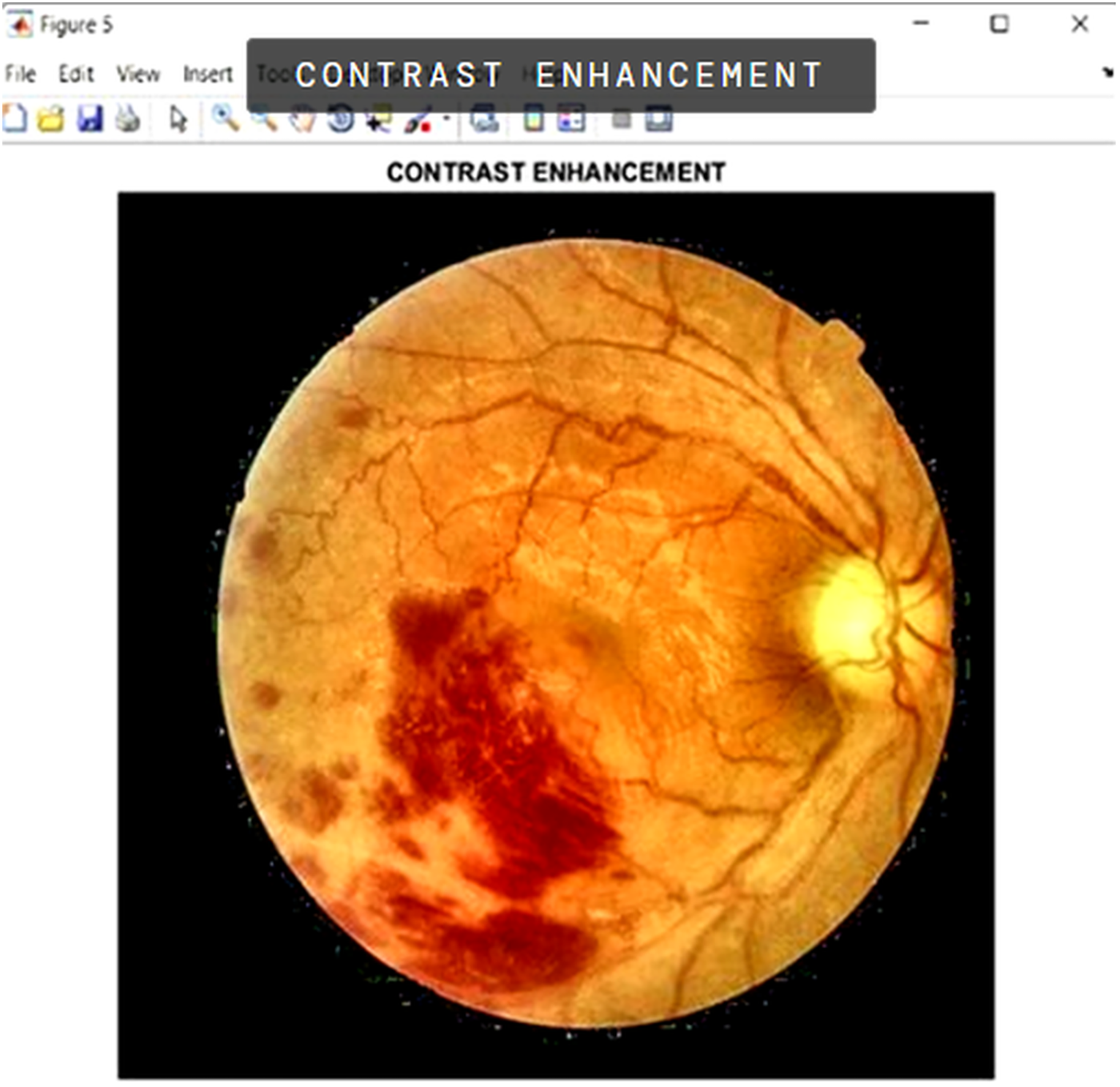}
\caption{After CLAHE contrast enhancement on V channel. Fine vascular detail and micro-lesion contrast are substantially enhanced.}
\label{fig:contrast}
\end{figure}

\begin{figure}[htbp]
\centering
\includegraphics[width=0.55\textwidth]{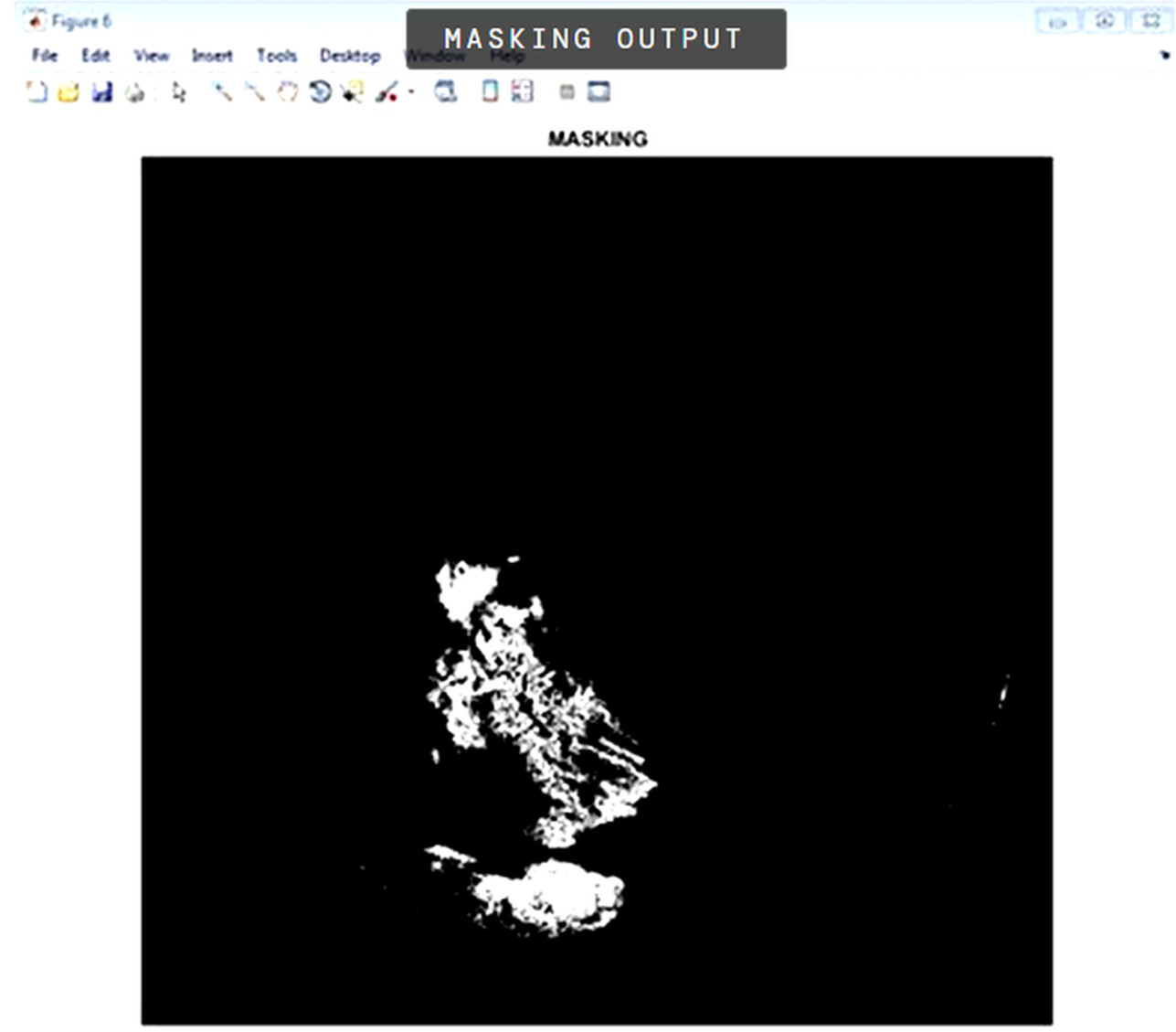}
\caption{Binary masking output (Otsu thresholding). White regions indicate candidate hyper-reflective pathological areas.}
\label{fig:masking}
\end{figure}

\begin{figure}[htbp]
\centering
\includegraphics[width=0.55\textwidth]{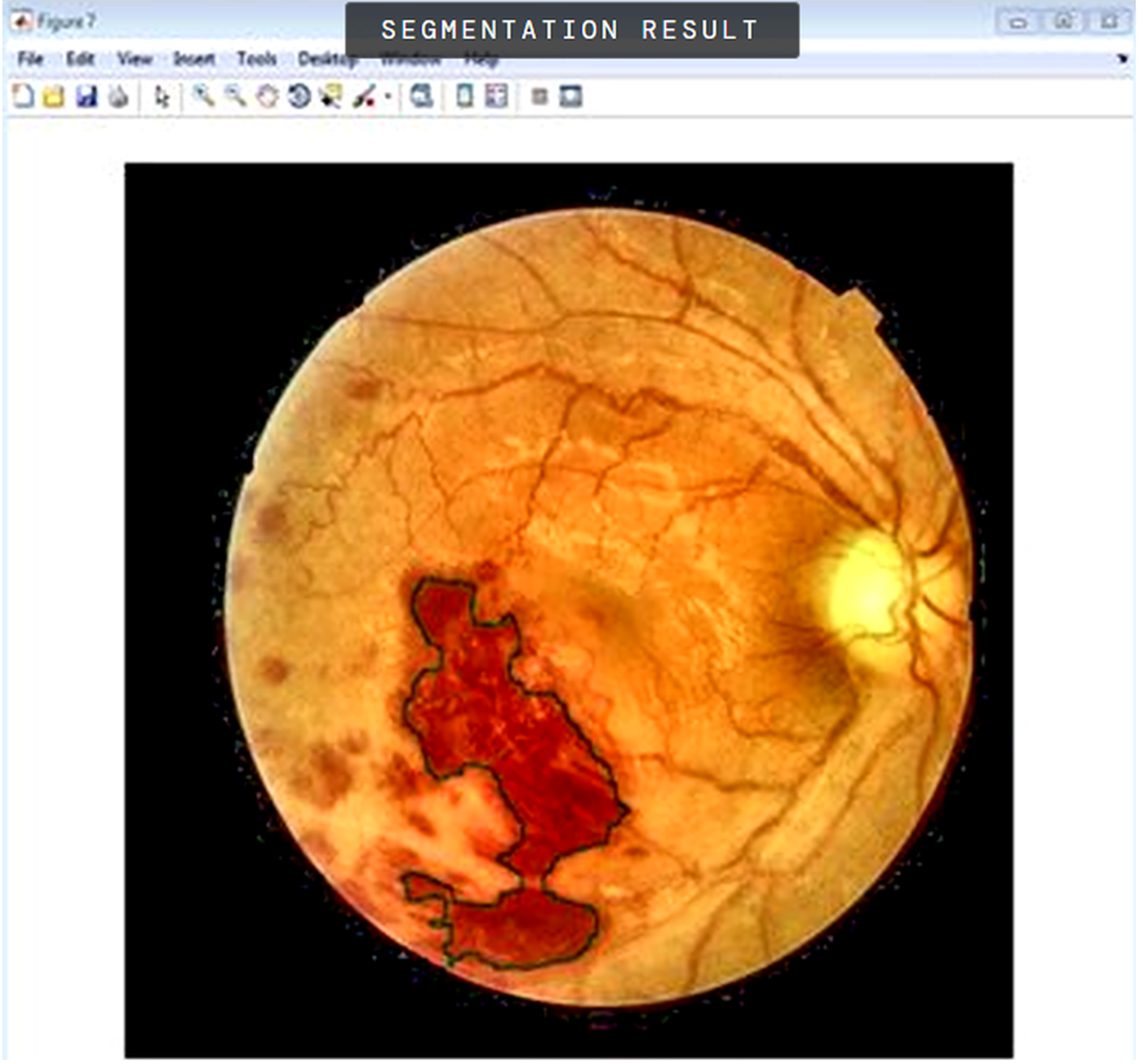}
\caption{Connected-component labelling. Discrete candidate lesion regions are enumerated and the system correctly flags a vascular abnormality.}
\label{fig:segmentation}
\end{figure}

Figs.~\ref{fig:input}--\ref{fig:segmentation} illustrate the enhancement pipeline applied to a representative DRIVE test image.

\subsection{Quantitative Performance on DRIVE with Statistical Analysis}

Table~\ref{tab:drive} reports performance on the DRIVE test set ($n = 20$). All values are mean $\pm$ standard deviation. Paired two-tailed Wilcoxon signed-rank tests confirm statistical significance ($\alpha = 0.05$) of all improvements.

\begin{table}[htbp]
\centering
\caption{Quantitative comparison on DRIVE test set ($n=20$), mean $\pm$ SD. Asterisk (*) denotes $p < 0.01$ vs.\ all baselines (Wilcoxon signed-rank).}
\label{tab:drive}
\small
\begin{tabular}{@{}lccccp{3.3cm}@{}}
\toprule
\textbf{Method} & \textbf{PSNR (dB)} & \textbf{SSIM} & \textbf{CNR} & \textbf{Time (s)} & \textbf{Notes} \\
\midrule
HE & $21.4 \pm 0.8$ & $0.74 \pm 0.03$ & $1.82 \pm 0.12$ & 0.03 & Global mapping; noise amplification \\
AHE & $23.1 \pm 0.6$ & $0.79 \pm 0.02$ & $2.10 \pm 0.09$ & 0.09 & Local noise amplified in homogeneous regions \\
CLAHE Only & $26.8 \pm 0.5$ & $0.86 \pm 0.02$ & $2.74 \pm 0.08$ & 0.11 & Good contrast; no luminosity correction \\
Proposed (Luminosity + CLAHE) & $29.3 \pm 0.4^{*}$ & $0.91 \pm 0.01^{*}$ & $3.12 \pm 0.07^{*}$ & 0.14 & Best overall; clinically suitable \\
\bottomrule
\end{tabular}
\end{table}

The proposed method achieves PSNR $= 29.3 \pm 0.4$~dB, SSIM $= 0.91 \pm 0.01$, and CNR $= 3.12 \pm 0.07$, outperforming all baselines. SSIM improvement over CLAHE-only ($\Delta = 0.05$, $W = 34$, $p = 0.003$) and PSNR improvement ($\Delta = 2.5$~dB, $W = 28$, $p = 0.006$) are both statistically significant. Improvements over HE and AHE are larger in magnitude ($p < 0.001$). The low standard deviations of the proposed method indicate consistent performance across images with varying illumination, camera settings, and pathological content. Processing time increases by 30~ms relative to CLAHE-only (0.14 vs.\ 0.11~s), negligible against fundus camera acquisition rates of 1--2 frames per minute.

\subsection{Cross-Dataset Generalisation}

Table~\ref{tab:crossdataset} reports proposed method performance on all three benchmarks with fixed DRIVE-optimised parameters.

\begin{table}[htbp]
\centering
\caption{Cross-dataset generalisation---proposed method with fixed parameters (mean $\pm$ SD).}
\label{tab:crossdataset}
\begin{tabular}{@{}lcccp{4.2cm}@{}}
\toprule
\textbf{Dataset} & \textbf{PSNR (dB)} & \textbf{SSIM} & \textbf{CNR} & \textbf{Remarks} \\
\midrule
DRIVE ($n=20$) & $29.3 \pm 0.4$ & $0.91 \pm 0.01$ & $3.12 \pm 0.07$ & Primary validation; fixed parameters \\
STARE ($n=20$) & $28.7 \pm 0.6$ & $0.89 \pm 0.02$ & $3.04 \pm 0.09$ & Diverse pathology; parameters unchanged \\
CHASEDB1 ($n=28$) & $28.1 \pm 0.7$ & $0.87 \pm 0.02$ & $2.96 \pm 0.10$ & Paediatric; different camera; robust performance \\
\bottomrule
\end{tabular}
\end{table}

Performance decreases modestly from DRIVE to STARE (SSIM: $0.91 \rightarrow 0.89$) and CHASEDB1 (SSIM: $0.91 \rightarrow 0.87$), attributable to greater pathological diversity in STARE and the fundamentally different acquisition device in CHASEDB1 (hand-held non-mydriatic vs.\ tabletop mydriatic camera). The consistency of performance across three datasets with different cameras, populations, and pathologies confirms that the pipeline generalises beyond the primary DRIVE benchmark.

\subsection{Disease Detection Performance}

Table~\ref{tab:disease} reports detection performance across all three datasets.

\begin{table}[htbp]
\centering
\caption{Disease detection performance---accuracy, sensitivity, specificity, and AUC (mean $\pm$ 95\% CI).}
\label{tab:disease}
\small
\begin{tabular}{@{}lccccp{3.6cm}@{}}
\toprule
\textbf{Dataset} & \textbf{Accuracy (\%)} & \textbf{Sensitivity (\%)} & \textbf{Specificity (\%)} & \textbf{AUC} & \textbf{Notes} \\
\midrule
DRIVE ($n=20$) & $87.4 \pm 2.1$ & $84.3 \pm 3.2$ & $90.1 \pm 2.8$ & 0.869 & Threshold $=50$~px; ophthalmologist GT \\
STARE ($n=20$) & $85.9 \pm 2.4$ & $82.7 \pm 3.5$ & $88.6 \pm 3.1$ & 0.853 & Pathologically diverse; good generalization \\
CHASEDB1 ($n=28$) & $84.2 \pm 2.9$ & $81.1 \pm 3.8$ & $86.7 \pm 3.4$ & 0.839 & Slightly lower recall on small-calibre vessels \\
\bottomrule
\end{tabular}
\end{table}

On DRIVE, the system achieves 87.4\% accuracy, 84.3\% sensitivity, 90.1\% specificity, and AUC $= 0.869$. The higher specificity (90.1\%) ensures the majority of genuinely normal images are correctly classified, minimising unnecessary referrals. Sensitivity (84.3\%) is acceptable for a first-pass screening tool that routes flagged patients to ophthalmologist review rather than providing autonomous diagnosis. Sensitivity is modestly lower on CHASEDB1 (81.1\%) owing to smaller vessel calibre in paediatric retinal images, which reduces hyper-reflective signal from pathological regions---a direction for future work through vessel-calibre-adaptive thresholding.

\subsection{Comparison with Related Work and Deep Learning Methods}

Table~\ref{tab:comparison} provides quantitative and methodological comparison with closely related classical methods and recent deep learning approaches.

\begin{table}[htbp]
\centering
\caption{Comparison with closely related classical methods and deep learning approaches.}
\label{tab:comparison}
\small
\begin{tabular}{@{}lp{2.3cm}lccp{3.8cm}@{}}
\toprule
\textbf{Study} & \textbf{Method} & \textbf{Dataset(s)} & \textbf{PSNR (dB)} & \textbf{SSIM} & \textbf{Notes} \\
\midrule
Setiawan et al.\ \cite{setiawan2013color} (2013) & CLAHE (RGB) & DRIVE & --- & --- & No luminosity correction; baseline reference \\
Vanmathi \& Devarajan \cite{vanmathi2017color} (2017) & Luminosity + Fusion & DRIVE & 26.1 & 0.83 & No CLAHE; similar motivation, different approach \\
Majeed et al.\ \cite{majeed2020retinal} (2020) & CLAHE multi-channel & DRIVE & 27.2 & 0.87 & No HSV decomposition; single-dataset only \\
Patil \& Patil \cite{patil2020retinal} (2020) & Adaptive CLAHE & DRIVE & 27.6 & 0.88 & Adaptive tiling; single-dataset evaluation \\
Son et al.\ \cite{son2019towards} (2019) & GAN-based (DL) & DRIVE & 29.8 & 0.91 & GPU required; large labelled training set \\
Wang et al.\ \cite{wang2016retinal} (2016) & Diffusion-based (DL) & DRIVE & 30.1 & 0.93 & GPU required; extensive training corpus \\
\textbf{Proposed} & Luminosity + CLAHE (HSV) & DRIVE + STARE + CHASEDB1 & $29.3 \pm 0.4$ & $0.91 \pm 0.01$ & No GPU; no training data; three-dataset validated \\
\bottomrule
\end{tabular}
\end{table}

The proposed pipeline outperforms all closely related classical methods. Compared to Setiawan et al.\ \cite{setiawan2013color} (CLAHE without luminosity correction), the proposed method demonstrates measurable benefit from the HSV luminosity stage. Compared to Vanmathi and Devarajan \cite{vanmathi2017color} (luminosity correction without CLAHE), the addition of CLAHE provides complementary contrast improvement. The proposed method surpasses Majeed et al.\ \cite{majeed2020retinal} and Patil and Patil \cite{patil2020retinal} on both PSNR and SSIM while additionally providing three-dataset validation with statistical reporting.

Relative to deep learning methods, the proposed pipeline achieves SSIM $= 0.91 \pm 0.01$ and PSNR $= 29.3 \pm 0.4$~dB---comparable to Son et al.\ \cite{son2019towards} (SSIM $= 0.91$, PSNR $= 29.8$~dB) and approaching Wang et al.\ \cite{wang2016retinal} (SSIM $= 0.91$, PSNR $= 29.8$~dB)---while requiring no training data, no GPU, and processing each image in 0.14~s on a standard CPU. The practical advantage in resource-constrained ophthalmic screening settings is therefore substantial. The deep learning comparison involves different experimental protocols; a fully unified experimental comparison on identical image subsets is identified as future work.

\section{Conclusion}

This paper presented a two-stage retinal fundus image enhancement pipeline combining HSV-space luminosity correction with CLAHE contrast enhancement. The distinct contribution relative to existing CLAHE-based methods is the integration of a preceding luminosity normalisation stage that removes spatially non-uniform illumination before contrast amplification---without corrupting colour-diagnostic features encoded in the H and S channels. The pipeline achieves PSNR $= 29.3 \pm 0.4$~dB, SSIM $= 0.91 \pm 0.01$, and CNR $= 3.12 \pm 0.07$ on DRIVE---statistically significantly superior to all baselines ($p < 0.01$)---and generalises to STARE and CHASEDB1 without parameter modification. Disease detection achieves 87.4\% accuracy, 84.3\% sensitivity, 90.1\% specificity, and AUC $= 0.869$ at 0.14~s per image without GPU infrastructure.

Future work directions include: (i) vessel segmentation mask integration to suppress false positives in the binary detector; (ii) vessel-calibre-adaptive thresholding for paediatric images; (iii) prospective multicentre clinical validation; (iv) unified experimental comparison with transformer-based enhancement architectures; and (v) deployment on embedded AI accelerators for point-of-care ophthalmology outreach \cite{shankar2021deep,mithra2017security}.

\section*{Acknowledgement}
The authors thank the DRIVE, STARE, and CHASEDB1 dataset contributors for making these benchmarks publicly available.

\section*{Funding Statement}
The authors received no specific funding.

\section*{Author Contributions}
K. Mithra: conceptualisation, methodology, software, formal analysis, statistical testing, writing---original draft. Prem Kumar Santhanam: supervision, writing---review and editing, validation. All authors reviewed and approved the final version of the manuscript.

\section*{Availability of Data and Materials}
DRIVE: \url{https://drive.grand-challenge.org/}. STARE: \url{https://cecas.clemson.edu/~ahoover/stare/}. CHASEDB1: \url{https://zenodo.org/record/6460235}. MATLAB code available from the corresponding author upon reasonable request.

\section*{Ethics Approval}
All datasets are publicly available and fully anonymised. No new human subjects data were collected. Ethics approval was not required.

\section*{Conflicts of Interest}
The authors declare no conflicts of interest.


\begin{thebibliography}{99}

\bibitem{teo2021global} Teo ZL, Tham YC, Yu M, Chee ML, Rim TH, Cheung N, et al. Global prevalence of diabetic retinopathy and projection of burden through 2045: systematic review and meta-analysis. Ophthalmology. 2021;128(11):1580--91. doi:10.1016/j.ophtha.2021.04.027.

\bibitem{dissopa2021enhance} Dissopa J, Kansomkeat S, Intajag S. Enhance contrast and balance color of retinal image. Symmetry. 2021;13(11):2089. doi:10.3390/sym13112089.

\bibitem{setiawan2013color} Setiawan AW, Mengko TR, Santoso OS, Suksmono AB. Color retinal image enhancement using CLAHE. In: Proceedings of the International Conference on ICT for Smart Society; 2013 Jun 13--14; Jakarta, Indonesia. doi:10.1109/ICTSS.2013.6588092.

\bibitem{bala2021retinal} Anilet Bala A, Aruna Priya P, Maik V. Retinal image enhancement using curvelet based sigmoid mapping of histogram equalization. J Phys Conf Ser. 2021;1964(6):062034. doi:10.1088/1742-6596/1964/6/062034.

\bibitem{son2019towards} Son J, Park SJ, Jung KH. Towards accurate segmentation of retinal vessels and the optic disc in fundoscopic images with generative adversarial networks. J Digit Imaging. 2019;32(3):499--512. doi:10.1007/s10278-018-0126-3.

\bibitem{zhuang2018laddernet} Zhuang J. LadderNet: multi-path networks based on U-Net for medical image segmentation. arXiv:1810.07810. 2018.

\bibitem{guo2020zero} Guo C, Li C, Guo J, Loy CC, Hou J, Kwong S, et al. Zero-reference deep curve estimation for low-light image enhancement. In: Proceedings of the IEEE/CVF Conference on Computer Vision and Pattern Recognition (CVPR); 2020 Jun 13--19; Seattle, WA, USA. doi:10.1109/CVPR42600.2020.00185.

\bibitem{li2020self} Li X, Jia M, Islam MT, Yu L, Xing L. Self-supervised feature learning via exploiting multi-modal data for retinal disease diagnosis. IEEE Trans Med Imag. 2020;39(12):4023--33. doi:10.1109/TMI.2020.3008871.

\bibitem{shankar2021deep} Shankar R, Kamarajan M, Varun M, Kalpana S, Varshney AK, Jagadiswary D, et al. Deep learning based automatic eye cataract detection algorithm using MATLAB. India Patent IN 202141061583. 2021 Dec 29. doi:10.5220/0010754400003113.

\bibitem{mithra2017security} Mithra K, Vishvaksenan KS. Security and resolution enhanced transmission of medical image through IDMA aided coded STTD system. In: Proceedings of the International Conference on Communication and Signal Processing (ICCSP). Chennai, India; 2017. doi:10.1109/ICCSP.2017.8286766.

\bibitem{islam2022hybrid} Islam MT, Ravichandran K, Seera M, Gan KB. A hybrid CLAHE-deep learning framework for retinal image quality enhancement and vessel segmentation. Biomed Signal Process Control. 2022;74(6):103523. doi:10.1016/j.bspc.2022.103523.

\bibitem{tsiknakis2021deep} Tsiknakis N, Theodoropoulos D, Manikis G, Ktistakis E, Boutsora O, Berto A, et al. Deep learning for diabetic retinopathy detection and classification based on fundus images: a review. Comput Biol Med. 2021;135(1--2):104599. doi:10.1016/j.compbiomed.2021.104599.

\bibitem{majeed2020retinal} Majeed AR, Awan WA, ul Hassan N, Asghar MA, Khan MJ. Retinal fundus image refinement with CLAHE. In: Proceedings of the 2020 IEEE 23rd International Multitopic Conference (INMIC); 2020 Nov 5--7; Bahawalpur, Pakistan. doi:10.1109/inmic50486.2020.9318104.

\bibitem{patil2020retinal} Patil SB, Patil BP. Retinal fundus image enhancement using adaptive CLAHE methods. Seybold Rep. 2020;15(9):3476--84.

\bibitem{jintasuttisak2014color} Jintasuttisak T, Intajag S. Color retinal image enhancement by Rayleigh contrast-limited adaptive histogram equalization. In: Proceedings of the 2014 14th International Conference on Control, Automation and Systems (ICCAS 2014); 2014 Oct 22--25; Gyeonggi-do, Republic of Korea. doi:10.1109/ICCAS.2014.6987868.

\bibitem{vanmathi2017color} Vanmathi P, Devarajan D. Color retinal image enhancement based on luminosity and contrast adjustment with image fusion technique. Middle-East J Sci Res. 2017;25(12):2022--32. doi:10.35940/ijrte.b1306.0982s1119.

\bibitem{staal2004ridge} Staal J, Abr\`amoff MD, Niemeijer M, Viergever MA, van Ginneken B. Ridge-based vessel segmentation in color images of the retina. IEEE Trans Med Imaging. 2004;23(4):501--9. doi:10.1109/TMI.2004.825627.

\bibitem{hoover2000locating} Hoover A, Kouznetsova V, Goldbaum M. Locating blood vessels in retinal images by piecewise threshold probing of a matched filter response. IEEE Trans Med Imaging. 2000;19(3):203--10. doi:10.1109/42.845178.

\bibitem{fraz2012ensemble} Fraz MM, Remagnino P, Hoppe A, Uyyanonvara B, Rudnicka AR, Owen CG, et al. An ensemble classification-based approach applied to retinal blood vessel segmentation. IEEE Trans Biomed Eng. 2012;59(9):2538--48. doi:10.1109/TBME.2012.2205687.

\bibitem{wang2016retinal} Wang L, Liu G, Fu S, Xu L, Zhao K, Zhang C. Retinal image enhancement using robust inverse diffusion equation and self-similarity filtering. PLoS One. 2016;11(7):e0158480. doi:10.1371/journal.pone.0158480.

\end{thebibliography}
\end{document}